\documentclass[a4paper,10pt]{article}
\usepackage{amsmath,amsfonts}
\usepackage{times}
\usepackage[T1]{fontenc}
\usepackage{epsfig}
\usepackage{amssymb}

\begin{document}

\title{\Huge{\bf A Skyrme-type proposal for baryonic matter}}

\author{C. Adam$^{a)}$\thanks{adam@fpaxp1.usc.es},
J. S\'{a}nchez-Guill\'{e}n
$^{a)b)}$\thanks{joaquin@fpaxp1.usc.es}, A. Wereszczy\'{n}ski
$^{c)}$\thanks{wereszczynski@th.if.uj.edu.pl}
       \\
       \\ $^{a)}$ Departamento de F\'isica de Part\'iculas, Universidad
       \\ de Santiago, and Instituto Galego de F\'isica de Altas Enerx\'ias
       \\ (IGFAE) E-15782 Santiago de Compostela, Spain
       \\
       \\ $^{b)}$ Sabbatical leave at: Departamento de F\'isica Te\'orica,
       \\ Universidad de Zaragoza, 50009 Zaragoza, Spain
       \\
       \\ $^{c)}$ Institute of Physics,  Jagiellonian University,
       \\ Reymonta 4, Krak\'{o}w, Poland}

\maketitle

\begin{abstract}
The Skyrme model is a low-energy effective field theory for QCD, where the
baryons emerge as soliton solutions. It is, however, not so easy within the
standard Skyrme model to reproduce the almost exact linear growth of the
nuclear masses with the baryon number (topological charge), due to the lack of
Bogomolny solutions in this model, which has also hindered analytical
progress. 
Here we identify a submodel within the
Skyrme-type low energy effective action which does have a Bogomolny bound and
exact Bogomolny solutions, and therefore, at least at the classical level, 
reproduces the nuclear masses by construction.
Due to its high
symmetry, this model qualitatively reproduces the main features of the liquid
droplet model of nuclei. 
Finally, we discuss under which circumstances the proposed sextic term, which 
is of an
essentially geometric and topological nature, can be expected to give a 
reasonable description of properties of nuclei. 
\end{abstract}
\section{Introduction}
The Skyrme model \cite{skyrme} is an effective low-energy action for QCD
\cite{witten},
where the primary ingredients are meson fields, whereas baryons appear
as solitonic excitations, and  the baryon
number is identified with the topological charge.
\\
The original Skyrme Lagrangian has the following form
\begin{equation}
L=L_2 + L_4  + L_0,
\end{equation}
where 
\begin{equation}
L_2=-\frac{f_{\pi}^2}{4} \; \mbox{Tr} \; (U^{\dagger} \partial_{\mu} U \; 
U^{\dagger} \partial^{\mu} U  )
\end{equation}
is the sigma model term, and
a quartic term, referred as Skyrme term, has to be added to
circumvent the standard Derrick argument for the non-existence of
static solutions,
\begin{equation}
L_4=-\frac{1}{32 e^2}\; \mbox{Tr} \; ([U^{\dagger} \partial_{\mu}
U,U^{\dagger} \partial_{\nu} U]^2 ).
\end{equation}
Here $U$ is a $2\times 2$ matrix-valued field with values in the group SU(2). 
The last term, which is optional from the point of view of the Derrick 
argument, is a potential
\begin{equation}
L_0= -\mu^2 V(U,U^{\dagger}),
\end{equation}
which explicitly breaks the the chiral symmetry. Its particular form is 
usually adjusted to a concrete physical
situation. The model has two constants, the pion decay constant
$f_{\pi}$ and the interaction parameter $e$. Additional
constants may appear from the potential.
\\
The modern point of view on the Skyrme model is to treat it as an
expansion in derivatives of the true non-perturbative low-energy
effective action of QCD, where higher terms in derivatives have
been neglected. However, as extended (solitonic) solutions have regions where
derivatives are not small, there is no reason for omitting such terms.
Therefore, one should take into account also
higher terms. In fact, many generalized Skyrme models have
been investigated \cite{modify marleau}, \cite{modify neto}, 
\cite{modify sk 2}, \cite{piette},
\begin{equation}
L=L_2 + L_4 + L_0+..., \label{skyrme full}
\end{equation}
where dots denote higher derivatives terms. 
A simple and natural extension of the Skyrme model is the addition of
sextic terms, among which one is rather special. Namely, we will
consider the expression
\begin{equation}
L_6=\frac{\lambda^2}{24^2} \; \left( \mbox{Tr}  \; (\epsilon^{\mu
\nu \rho \sigma} U^{\dagger} \partial_{\mu} U \; U^{\dagger}
\partial_{\nu} U \; U^{\dagger} \partial_{\rho} U) \right)^2. \label{6}
\end{equation}
In standard phenomenology, the addition of this term to the effective action 
represents the inclusion of the interactions generated by the vector mesons 
$\omega$.   In fact, this term effectively appears if one considers a 
massive vector field coupled to the chiral field via the baryon density 
\cite{modify sk 1}. Further, this term is at most quadratic in time
derivatives (like the quartic Skyrme term) and allows for a standard time
dynamics and hamiltonian formulation. In addition, it leads to a significant 
improvement in the Skyrme model phenomenology when applied to nucleons.
Indeed, as explained first in \cite{modify sk 2}, once the sextic
term is present, it becomes the main responsible for stabilization,
and then the quartic contribution changes sign as it corresponds to
the scalar exchange it represents (solving an old puzzle). This
compensation with the quadratic term holds also for the moments of
inertia, when the rotation of all the mass is taken into account, as
it should in the classical computation. 
\\
In this letter we want to study the model restricted to the potential 
and the sextic term, $L_{06} = L_6 + L_0$, because this submodel has some unique
properties. First of all, it has a huge amount of symmetry \cite{ab-dif}
and, therefore,
it is integrable in the sense of generalized integrability \cite{alvarez}
(its symmetries and integrability properties shall be discussed in detail
in a separate publication; its symmetries are also important for its rather
close relation to the liquid droplet model of nuclei, as we shall discuss 
at length in
the last section). As a consequence, the model has infinitely many
exact solutions in all topological sectors, such that both energies and
profiles can be determined exactly. Finally, the model has a Bogomolny bound
which is saturated by all the exact solutions we construct below. 
The existence of static solutions which saturate a Bogomolny bound is very 
welcome for the description of nuclei, for the following reasons. Firstly, 
the resulting soliton energies obey an exactly linear relation with the 
baryon charges. Physical nuclei are well-known to obey this linear law with 
a rather high precision. Secondly, binding energies of higher solitons are 
zero, again as a consequence of their Bogomolny nature. This conforms rather 
well with the binding energies of physical nuclei, which are usually quite 
small (below the 1\% level). Thirdly, the forces between sufficiently 
separated solitons are exactly zero. This result is a consequence of 
another crucial feature of our solitons, namely their compact nature. 
Again, this absence of interactions, although not exactly true, is a 
rather reasonable approximation for physical nuclei, given the very 
short range character of interactions between them.  
\\
So we find a rather striking coincidence between some qualitative features 
of nuclei, on the one hand, and properties of our {\em classical} soliton 
solutions, on the other hand. One important question is, of course, whether 
this coincidence can be maintained at the quantum level. A detailed 
investigation of the quantization of the model is beyond the scope of this 
letter, but we shall comment further on it in the discussion section. 
In any case, the model seems to correspond to a rather non-trivial "lowest 
order" effective field theory approximation to nuclei which already 
reproduces some of their features quite well. 
We also want to remark that part of the pseudoscalar
meson dynamics is possibly taken into account already by the potential $L_0$, 
which
breaks the chiral symmetry, as goldstone condensation.
\\
 All the unique properties of the model may be ultimately
traced back to the geometric properties of the proposed term $L_6$, 
i.e., to the fact that it is 
the square of the pullback of the volume form on the target space three-sphere
$S^3$ (we remind that as a manifold SU(2) $\simeq S^3$) or, equivalently, 
the square of the topological (baryon) current.
We remark that models which are similar in some aspects, although with a 
different target space geometry,  have been
studied in \cite{Tchr 1}, \cite{Tchr 2}.
Further, the model studied in this letter, as well as its "baby Skyrme" 
version in 2+1 dimensions have already been introduced in  \cite{Tchr 3}. 
There, 
the main aim was a study of more general properties of Skyrme models in 
any dimension. Concretely,
the limiting behaviour of the full generalized Skyrme model for small 
couplings of the quadratic and quartic terms $L_2$ and $L_4$ was studied 
numerically. In addition, an exact solution for the simplest hedgehog 
ansatz was constructed, both in 2 and in 3 dimensions. For the 
three-dimensional solution, a rather complicated potential was chosen in 
order to have exponentially localized solutions, whereas in this letter we 
shall focus on the case of the simple standard Skyrme potential, which 
naturally leads to compact solitons. Besides, 
our main purpose is to make contact with the phenomenology of nuclei.
The 2+1 dimensional baby Skyrme version of the model has been further 
investigated in 
\cite{GP} and recently in \cite{restricted-bS}, with results which are 
qualitatively similar to the ones we shall find in the sequel (e.g. compact 
solitons, infinitely many symmetries, Bogomolny bounds).

\section{Exact solutions}
\begin{figure}[h!]
\includegraphics[angle=0,width=0.55 \textwidth]{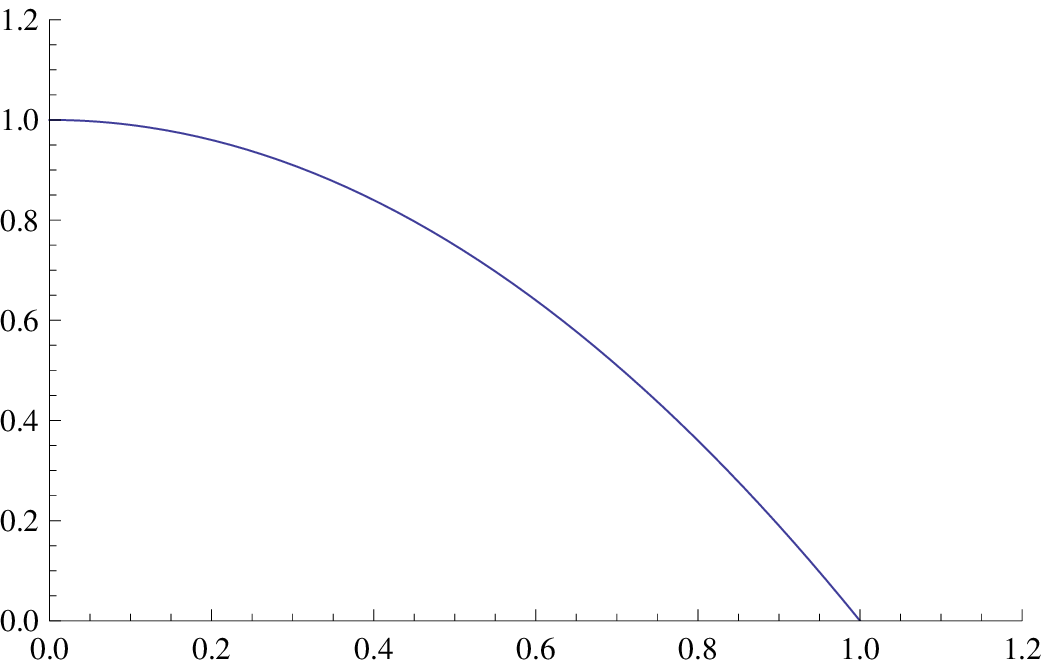}
\includegraphics[angle=0,width=0.55 \textwidth]{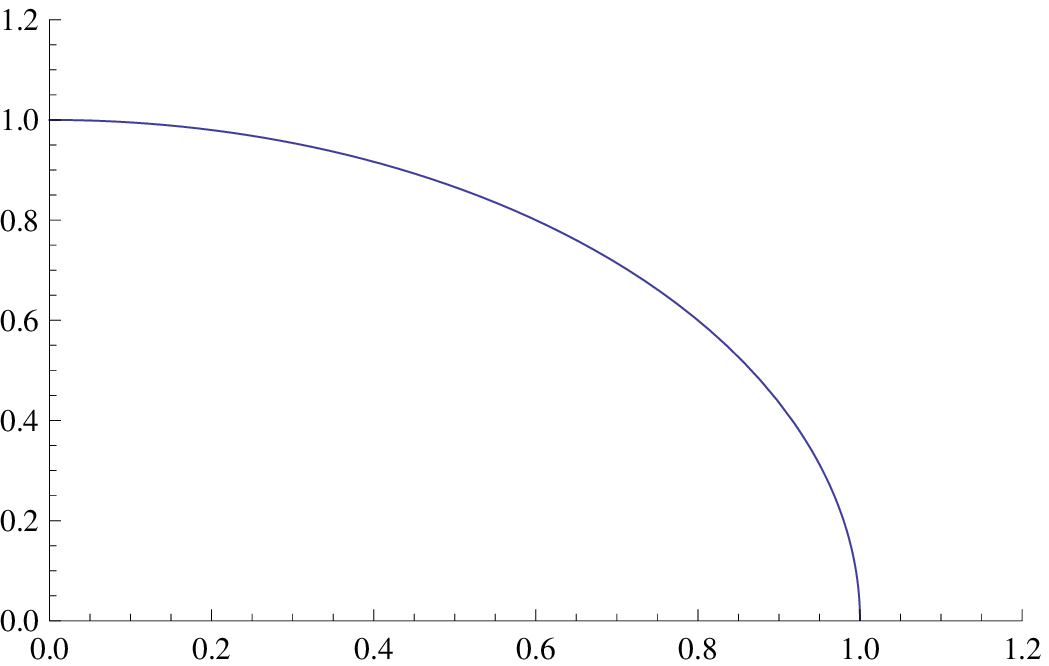}
\caption{Normalized energy density (left figure) and topological charge 
density (right figure) as a function of the
rescaled radius $\tilde r$, for topological charge n=1. For $|n|>1$, the
height of the densities remains the same, whereas their radius grows like
$|n|^\frac{1}{3}$  }\label{rys1}
\end{figure}

The lagrangian of the proposed restriction of the Skyrme model is
\begin{equation}
L_{06}=\frac{\lambda^2}{24^2 } \;  \left( \mbox{Tr}  \; ( \epsilon^{\mu \nu 
\rho \sigma} U^{\dagger} \partial_{\mu} U \;
U^{\dagger} \partial_{\nu} U \;
U^{\dagger} \partial_{\rho} U) \right) ^2 - \mu^2 V(U,U^{\dagger}).
\end{equation}
We start from 
the standard parametrization for $U$ by a real scalar field
$\xi$ and a three component unit vector $\vec{n}$ ($\vec \tau$ are the Pauli
matrices), 
$$
U=e^{i \xi \vec{n} \cdot \vec{\tau}}.
$$
The vector field may be related to a complex scalar $u$ by the 
stereographic projection
$$
\vec{n}=\frac{1}{1+|u|^2} \left( u+\bar{u}, -i ( u-\bar{u}),
|u|^2-1 \right)
$$
giving finally ($\tau_\pm = (1/2)(\tau_1 \pm i \tau_2) $)
$$ U^{\dagger} \partial_{\mu} U=
W^\dagger \left( -i\xi_{\mu} \tau_3+\frac{2\sin \xi}{1+|u|^2} 
\left( e^{i\xi} u_{\mu} \tau_+-
e^{-i\xi} \bar{u}_{\mu} \tau_-\right) \right) W
$$
where the SU(2) matrix field $W$ is 
$$
W=  (1+u\bar u)^{- \frac{1}{2}} \left(
\begin{array}{cc}
1 & iu \\
i\bar u & 1 
\end{array} \right) 
$$
and obviously cancels in the lagrangian.
Using this parametrization we get  ($u_\mu \equiv \partial_\mu u$, etc.)    
\begin{equation}
L_{06}= -\frac{  \lambda^2 \sin^4 \xi}{(1+|u|^2)^4} \;\left(  \epsilon^{\mu 
\nu \rho \sigma} \xi_{\nu} u_{\rho} \bar{u}_{\sigma} \right)^2
-\mu^2 V(\xi)
\end{equation}
where we also assumed that the potential only depends on $\mbox{tr} \, U$.
The Euler--Lagrange equations read ($V_\xi \equiv \partial_\xi V$)
$$ \frac{\lambda^2 \sin^2 \xi}{(1+|u|^2)^4} \partial_{\mu} ( \sin^2 \xi \; 
H^{\mu}) - \mu^2 V_{\xi}=0,$$
$$ \partial_{\mu} \left( \frac{K^{\mu}}{(1+|u|^2)^2} \right)=0,$$
where 
$$ H_{\mu} = \frac{\partial (  \epsilon^{\alpha \nu \rho \sigma} \xi_{\nu} 
u_{\rho} \bar{u}_{\sigma})^2}{ \partial \xi^{\mu}}, \;\;\; K_{\mu} = 
\frac{\partial (  \epsilon^{\alpha \nu \rho \sigma} \xi_{\nu} u_{\rho} 
\bar{u}_{\sigma})^2}{\partial \bar{u}^{\mu}}.$$
These objects obey the useful formulas 
$$
H_{\mu} u^{\mu}=H_{\mu} \bar{u}^{\mu}=0, \; K_{\mu}\xi^{\mu}=K_{\mu}
u^{\mu}=0, 
\;\; H_{\mu} \xi^{\mu}=K_{\mu} \bar{u}^{\mu} = 2 (  \epsilon^{\alpha \nu \rho 
\sigma} \xi_{\nu} u_{\rho} \bar{u}_{\sigma})^2. 
$$ 
We are interested in static topologically non-trivial solutions. Thus $u$ must 
cover the whole complex plane ($\vec{n}$ covers at least once $S^2$) 
and $\xi \in [0,\pi]$. The natural (hedgehog) ansatz is
$$ \xi = \xi (r), \;\;\; u(\theta, \phi) = g (\theta) e^{in \phi}.$$  
Then, the field equation for $u$ reads
$$ 
\frac{1}{\sin \theta} \partial_{\theta} \left( \frac{ g^2g_\theta}{(1+g^2)^2 
\sin \theta} \right) - \frac{gg_\theta^2}{(1+g^2)^2\sin^2 \theta}=0,
$$
and the solution with the right boundary condition is
$$ 
g(\theta) = \tan \frac{\theta}{2}.
$$
Observe that this solution holds for all values of $n$. 
The equation for the real scalar field is 
$$
\frac{n^2\lambda^2 \sin^2 \xi }{2r^2} \partial_r \left(\frac{\sin^2 \xi \; 
\xi_r}{r^2} \right) - \mu^2 V_{\xi}=0.
$$ 
This equation can be simplified by introducing the new variable 
$z=\frac{\sqrt{2}\mu r^3}{3 |n|\lambda}$, 
\begin{equation} \label{xi-eq}
\sin^2 \xi \; \partial_z \left(\sin^2 \xi \; \xi_z\right) -  V_{\xi}=0,
\end{equation}
and may be integrated to 
\begin{equation} 
 \frac{1}{2} \sin^4 \xi \; \xi^2_z=V, \label{bps eq}
\end{equation}
where we chose vanishing integration constant to get finite energy solutions. 
Now, we have to specify a concrete potential. 
The most obvious choice is the standard Skyrme potential 
\begin{equation}
V=\frac{1}{2}\mbox{Tr} (1-U) \;\; \rightarrow \;\; V(\xi)=1- \cos \xi. 
\end{equation}
Thus, 
$$  
\sin^2 \xi \; \xi_z=\pm \sqrt{2(1-\cos \xi)}\;\; \Rightarrow \;\; 
\int \frac{\sin^2 \xi}{\sin \xi /2}= \pm 2(z-z_0).
$$
The general solution reads 
$$ 
\cos^3 \frac{\xi}{2} = \pm \frac{3}{4} (z-z_0).
$$
Imposing the boundary conditions for
topologically non-trivial solutions we get 
\begin{equation}
\xi = \left\{
\begin{array}{lc}
2 \arccos \sqrt[3]{ \frac{3z}{4} } & z \in \left[0,\frac{4}{3} \right] \\
0 & z \geq \frac{4}{3}.
\end{array} \right.
\end{equation}
The corresponding energy is
\begin{equation}
E=\int d^3x \left( -\frac{\lambda^2 \sin^4 \xi}{(1+|u|^2)^4} 
(\nabla_r \xi )^2 ( \nabla_{\theta} u \nabla_{\phi} \bar{u} - \nabla_{\phi} u 
\nabla_{\theta} \bar{u})^2 +\mu^2 V  \right).
\end{equation}
Inserting the solution for $u$ and (\ref{bps eq}) we find 
\begin{eqnarray}
E &=& 4\pi \int r^2 dr \left( \frac{\lambda^2 n^2\sin^4 \xi}{4r^4} \xi^2_r 
+\mu^2 V  \right) \nonumber \\
&=& 4\pi \cdot 2\mu^2 \int r^2 dr V(\xi (r))= 4 \sqrt{2}\pi 
\mu \lambda |n| \int dz V (\xi (z))  \nonumber \\
&=& 8 \sqrt{2}\pi \mu \lambda |n| \int_0^{4/3} 
\left(1-\left( \frac{3}{4} \right)^{2/3} z^{2/3}\right)dz = 
\frac{64\sqrt{2} \pi}{15} \mu \lambda |n| .
\end{eqnarray}
 The solution is of 
the compacton type, i.e., it has a finite support 
(compact solutions of a similar
type in different versions of the
baby Skyrme models have been found in \cite{GP},
\cite{comp-bS}).
The function $\xi$ is continuous 
but its first derivative is not. The jump of the derivative is, in fact, 
infinite at the compacton boundary
$z=4/3$, as the left derivative at this point tends to minus infinity. 
Nevertheless, the energy density and the topological charge density
(baryon number density) are continuous functions at the compacton boundary,
and the field equation (\ref{xi-eq}) is well-defined there, so the solution is
a strong solution. The reason is that $\xi_z$ always appears in the
combination $\sin^2 \xi \, \xi_z$, and this expression is finite (in fact, zero)
at the compacton boundary. We could make the discontinuity disappear
altogether by introducing a new variable $\tilde \xi$ instead of $\xi$
which satisfies
$
\tilde \xi_z = \sin^2 \xi \, \xi_z .
$
We prefer to work with $\xi$ just because this is the standard
variable in the Skyrme model.
\\
In order to extract the energy density it is useful to rewrite the energy with
the help of the rescaled radial coordinate
\begin{equation}
\tilde r = \left( \frac{\sqrt{2}\mu}{4 \lambda} \right)^\frac{1}{3} r =
\left(\frac{3|n|z}{4}\right)^\frac{1}{3}
\end{equation}
like
$$
E = 8 \sqrt{2}  \mu \lambda \left( 4 \pi \int_0^{|n|^\frac{1}{3}} d\tilde r 
\tilde r^2 (1- |n|^{-\frac{2}{3}}\tilde r^2) \right) 
$$
such that the energy density per unit volume (with the unit of length set by
$\tilde r$) is
\begin{equation}
{\cal E}=   8 \sqrt{2} \mu \lambda (1- |n|^{-\frac{2}{3}} \tilde r^2 ) . 
\end{equation}
In the same fashion we get for the topological charge (baryon number), see
e.g. chapter 1.4 of \cite{mak}
\begin{eqnarray} \label{Bcharge}
B &=& -\frac{1}{\pi^2} \int d^3 x \frac{\sin^2 \xi }{(1+|u|^2)^2}
i\epsilon^{mnl} \xi_m u_n \bar u_l \\
&=& \frac{2n}{\pi} \int dr \sin^2 \xi \, \xi_r =
\frac{4n}{\pi} \int_0^\frac{4}{3} dz
\left( 1-\left(\frac{3}{4}\right)^\frac{2}{3} z^\frac{2}{3}\right)^\frac{1}{2}
\nonumber \\
&=& \mbox{sign} (n) \frac{4}{\pi^2} \left(4\pi \int_0^{|n|^\frac{1}{3}} 
d\tilde r \tilde
r^2 (1- |n|^{-\frac{2}{3}}\tilde r^2)^\frac{1}{2} \right) =n \nonumber
\end{eqnarray}
and for the topological charge density per unit volume
\begin{equation}
{\cal B} =  \mbox{sign} (n) \frac{4}{\pi^2} (1- |n|^{-\frac{2}{3}}\tilde
r^2)^\frac{1}{2} .
\end{equation}
Both densities are, of course, zero outside the compacton radius $\tilde r
=|n|^\frac{1}{3}$. 
We remark that the values of the densities at the center $\tilde r=0$ are
independent of the topological charge $B=n$, whereas the radii grow like
$n^\frac{1}{3}$.  For $n=1$, we plot the two densities in Fig. (\ref{rys1}),
where we normalize both densities (i.e., multiply them by a constant) such
that their value at the center is one.
\\
We now want to compare the phenomenological parameters of our model 
(masses and radii) to the corresponding values for physical nuclei.
One should keep in mind, of course, that the comparison is done at the purely classical level, and all quantum corrections are absent.
First, observe that the energy of the solitons is proportional to the 
topological (baryon) charge
$$ E=E_0 |B|, $$
where $E_0=64\sqrt{2}\pi \mu\lambda / 15$. Such a linear dependence is a basic
feature in nuclear physics.  Let us fix the energy scale by setting
$E_0 = 931.75 $ MeV. This is equivalent to the assumption that the mass 
of the $B=4$ solution is equal to the mass of He$^4$, which is usually
assumed because the ground state of He$^4$ has zero spin and isospin. 
Therefore, corrections to the mass from spin-isospin interactions 
are absent. In table (\ref{table}) we compare the energies of the solitons 
in our model with the experimental values.
We find that 
the maximal deviation in our model is only about $0.7\%$.
For the numerical determination of soliton masses in current versions of the
Skyrme model we refer to \cite{massive skyrme} (the standard massive Skyrme
model) and to \cite{vec skyrme} (the vector Skyrme model, where a coupling of
the Skyrme field to vector mesons is used instead of the quartic Skyrme term
for the stabilisation of the Skyrmions). There, typically, the Skyrmions with
low baryon number are heavier by a few percent, whereas they reproduce the
linear growth of mass with baryon number for higher baryon number. In
\cite{kope} the Skyrmion masses have been determined with the help of the
rational map approximation \cite{rationalmap} for Skyrmions, with similar
results.     
\\
Secondly, the sizes of the solitons can be easily computed and read 
$$ R_B= R_0 \sqrt[3]{|B|}, \;\;\; R_0=\left( \frac{2\sqrt{2} \lambda }{\mu} 
\right)^{\frac{1}{3}},$$
which again reproduces the well-known experimental relation. 
The numerical value is fixed by assuming  $R_0=1.25$ fm.

\begin{table}
\begin{center}
\begin{tabular}{|c|c|c|}
\hline
B & $E_{06}$ &   $E_{experiment}$ \\
\hline 
1 &  931.75 &  939 \\
2 &  1863.5 &  1876 \\
3 &  2795.25  &  2809 \\
4 &  3727 &  3727 \\
6 &  5590.5 &  5601 \\
8 &  7454 &  7455 \\
10 &  9317.5 &  9327 \\
\hline
\end{tabular}  
\caption{Energies of the soliton solutions in our model ($E_{06}$), 
compared with 
the experimental masses of physical nuclei. 
All numbers are in MeV} \label{table}
\end{center}
\end{table}

\section{Bogomolny bound}
Now, we show that our solitons are of the BPS type and saturate a Bogomolny
bound. 
Let us mention here that a Bogomolny bound also exists for the original Skyrme
model $L_2 + L_4$, but it is easy to prove that non-trivial solutions of this
model cannot saturate the bound (see, e.g., \cite{mak}; this bound has been
found already by Skyrme himself \cite{skyrme}).
\\
The energy functional reads
$$
E=\int d^3 x \left(  \frac{\lambda^2 \sin^4 \xi}{(1+|u|^2)^4} 
(\epsilon^{mnl} i \xi_m u_n\bar{u}_l)^2 +\mu^2 V \right) = 
$$ 
$$ 
= \int d^3 x \left( \frac{\lambda \sin^2 \xi}{(1+|u|^2)^2} 
\epsilon^{mnl} i \xi_mu_n\bar{u}_l \pm \mu \sqrt{V} \right)^2 \mp \int d^3 x 
\frac{2\mu \lambda \sin^2 \xi \sqrt{V}}{(1+|u|^2)^2} \epsilon^{mnl} i \xi_m 
u_n \bar{u}_l 
$$
$$ 
\geq  \mp \int d^3 x \frac{2\mu \lambda \sin^2 \xi \sqrt{V}}{(1+|u|^2)^2} 
\epsilon^{mnl} i \xi_m u_n \bar{u}_l =
$$
\begin{equation} \label{bobo}
\pm (2\lambda \mu \pi^2 )\left[ \frac{-i}{\pi^2}
\int d^3 x \frac{ \sin^2 \xi \sqrt{V}}{(1+|u|^2)^2} 
\epsilon^{mnl}  \xi_m u_n \bar{u}_l \right] 
\equiv 2\lambda \mu \pi^2 C_1 |B|
\end{equation}
where $B$ is the baryon number (topological charge) and the sign has to
be chosen appropriately (upper sign for $B>0$). If we replace $\sqrt{V}$ by
one, then the result (i.e., the last equality in (\ref{bobo}))
follows immediately (and the constant $C_1=1$). Indeed, for $V=1$
the expression in brackets is just the topological charge (\ref{Bcharge}).
An equivalent derivation, which shall be
useful below, starts with the observation that this expression is just the
base space integral of the pullback of the volume form on the
target space $S^3$, normalized to one. Further, while the target space $S^3$
is covered once, the base space $S^3$ is covered $B$ times, which 
implies the result. 
The same argument continues to hold with the factor $\sqrt{V}$ present 
(remember
that $V=V(\xi)$), up to a constant $C_1$. Indeed, we just have to introduce
a new target space coordinate $\bar\xi $ such that
\begin{equation} \label{xi-xi'}
\sin^2 \xi \sqrt{V(\xi)} \, d\xi = C_1  \sin^2 \bar \xi  \, d \bar\xi  .
\end{equation}
The constant $C_1$ and a second constant $C_2$, which is provided by
the integration of Eq. (\ref{xi-xi'}), are needed to impose the two conditions
$\bar\xi (\xi=0)=0$ and $\bar \xi (\xi =\pi) =\pi$, which have to hold if 
$\bar\xi$ is a
good coordinate on the target space $S^3$. Obviously, $C_1$ depends on the
potential $V(\xi)$. Specifically, for the standard Skyrme potential
$V=1-\cos\xi$, $C_1 $ is
$$
C_1 = \frac{32\sqrt{2}}{15\pi}
$$
as may be checked easily by an elementary integration. We remark that an
analogous Bogomolny bound in one lower dimension has been derived
in \cite{deI-W}, \cite{restricted-bS} for the baby Skyrme model.
\\
The Bogomolny inequality is saturated by configurations obeying the first order 
Bogomolny equation
$$ \frac{\lambda \sin^2 \xi}{(1+|u|^2)^2} \epsilon^{mnl} i \xi_mu_n\bar{u}_l 
= \mp \mu \sqrt{V} ,$$
which, in the case of our ansatz, reduces to the square root of equation
(\ref{bps eq}).  The saturation of the energy-charge inequality by our 
solutions proves their stability. It is not possible to find configurations
with lesser energy in a sector with a fixed value of the 
baryon charge. 
\section{Discussion}
In this letter we proposed an integrable limit of the full Skyrme model which 
consists of two terms: the square of the pullback of the target space volume 
(topological density) and a non-derivative part, i.e., a potential. 
Both terms are needed to circumvent the Derrick argument.
Then we explicitly solved the static model for a specific choice of the 
potential (the standard Skyrme potential). The resulting solitons satisfy, in
fact, a Bogomolny equation.
These exact Bogomolny solutions provide a linear relation between
soliton energy (=nuclear mass) and topological charge (=baryon number $B$),
reproducing the experimental nuclear masses with a high precision.  
Besides, these solitons have the remarkable property of being compact, which
allows to define a strict value of the soliton size (=nuclear radius). 
The resulting radii $R$, too, 
follow the standard experimental relation $R\sim |B|^{1/3}$
with a high precision. 
\\
These findings lead to the question of the nature and quality of the
approximation which our model provides for the properties of physical nuclei.
Obviously, the model as it stands cannot reproduce all features of nuclei,
even qualitatively,
because some essential ingredients are still missing. 
First of all, the binding energy of higher nuclei is zero due to the Bogomolny
nature of the solutions. Although not entirely correct for physical nuclei,
this is, however, not such a bad approximation, because the nuclear binding
energies are known to be rather small. Their smallness is, in fact, one of the
motivations for the search for theories which saturate a Bogomolny bound.
Secondly, there are 
no pionic excitations, because the corresponding term in the Lagrangian is
absent. This is related to the complete absence of forces between separated,
non-overlapping solitons. The absence of forces is a direct consequence of the
compact nature of these solitons, because
several non-overlapping solitons still represent an exact solution of the
field equations. 
Physical nuclei are not strictly
non-interacting, but given the very short range character of forces between
nuclei, the absence of interactions in the model may, in fact, 
be welcome from a phenomenological
point of view, within a certain approximation. Further, for physical nuclei a
finite radius may be defined with good accuracy, so the compact nature of 
the solitons may be a virtue also from this point of view. 
\\
For the energy density we find that it is of the core type
(i.e., larger in the center and decreasing towards the boundary), see Fig. 1.
The baryon density profile is again of the core type, but flatter near the
center, and with a smaller and more pronounced surface (=region where the
density decreases significantly). 
For physical nuclei the density profile is quite flat (almost constant) and for
some nuclei even with a shallow valley in the center, so here the
phenomenological coincidence is reasonable but not perfect. Let us also
mention that the independence of the profile heights of the baryon number
conforms well with the known properties of nuclei.
\\
Our results for the profiles, 
however, have to
be taken with some care. First of all, they depend on the form of the
potential term, in contrast to the linear mass-charge relation (which holds
for all potentials) or the compact nature of the solitons
(which holds for a wide class of
potentials). The second argument is related to the huge amount of symmetry of
the model. Indeed, for the energy functional for static field configurations,
the volume-preserving diffeomorphisms on the
three-dimensional base space are a subset of these symmetries. In
physical terms, all deformations of solitons
which correspond to these volume-preserving
diffeomorphisms may be performed without any cost in energy. 
But these deformations are exactly
the allowed deformations for an ideal, incompressible droplet of liquid where
surface contributions to the energy are neglected. 
These symmetries are not symmetries of a physical nucleus. A physical nucleus
has a definite shape, and deformations which change this shape cost energy.
Nevertheless, deformations which respect the local volume conservation (i.e.,
deformations of an ideal incompressible liquid) cost much less energy
than volume-changing deformations, as an immediate consequence 
of the liquid droplet model of nuclear matter. 
\\
This last observation also further explains the nature of the approximation our
model provides for physical nuclei. It reproduces some of the classical
features of the
nuclear liquid droplet model at least on a qualitative level, and the huge
amount of symmetries of the model is crucial for this fact.  
Its soliton energies, e.g., correspond to the bulk (volume) contribution of
the liquid droplet model, with the additional feature that the energies are
quantized in terms of a topological charge.   
\\  
In other words, the model provides,
besides a conceptual understanding with exact solutions, 
a new starting point or ``zero order''
approximation which is different from other approximations. It already covers
some nuclear droplet properties of nuclear matter, and is topological in
nature. For a more quantitative and phenomenological application to nuclei,
obviously both the inclusion of additional terms and the quantization of some
degrees of freedom are necessary.   
\\
So let us briefly discuss the question of possible generalizations of the
model. A first,
simple generalization consists in the choice of different potentials. The
resulting solitons continue to saturate a Bogomolny bound, therefore the
linear relation $E\sim |B|$ between energy and baryon number continues to
hold. The energy and baryon charge densities for a spherically symmetric
ansatz (hedgehog), and even the compact or
non-compact nature of the solitons, however, will depend
on the specific form of the potential.  
\\
A further generalization consists in including additional terms in
the Lagrangian (like the terms $L_2$ and $L_4$ of the standard Skyrme model)
which we have neglected so far. 
From the effective field theory point of view
there is no reason not to include them. 
If we omit, e.g., the kinetic term for the
chiral fields, then   
there are no obvious pseudo-scalar degrees of freedom ($\eta, \vec{\pi}$). 
These additional terms break the huge symmetry of the original model, such
that the solitons now have fixed shapes. In order to describe nuclei, these
shapes should be at least approximately spherically symmetric. A detailed
investigation of this issue is beyond the scope of the present letter, but
let us mention that at least under simple volume-preserving deformations from
a spherical to an ellipsoidal shape both the $E_2$ term and the $E_4$ term
energetically prefer the spherical shape.
Further, the reasonable qualitative success of the restricted model might 
indicate that the additional terms should be small in some sense (e.g., their
contribution to the total energy should not be too big). This opens the 
possibility
of an approximate treatment, where the solitons of the restricted model
$L_{06}$ provide the solutions to ``zeroth order'' (with all the topology and
reasonable energies already present), whereas the additional terms provide
corrections, which may be adapted to the experimental energies and
shapes of nuclei.
\\
Further, a more realistic treatment certainly requires the investigation of 
the issue of quantization. We emphasize again that the rather good 
phenomenological properties of
the model up to now are based exclusively on the classical solutions, and it 
is a different question whether quantum corrections are sufficiently small or 
well-behaved such that this success carries over to the quantized model. A 
first step in this direction consists in applying the rigid rotor quantization 
to the (iso-) rotational degrees of freedom, 
as has been done already for the
standard Skyrme theory \cite{skyrme-quant-wit}, for some recent applications
to the spectroscopy of nuclei see e.g. \cite{skyrme-quant-sut}. 
Some first calculations related to this rigid rotor quantization have 
been done already, with encouraging results. A second issue is, of course, 
the collective coordinate quantization of the (infinitely many) remaining 
symmetries. This point certainly requires further study. A pragmatic 
approach could assume that a more realistic application to nuclei requires, 
in any case,  the inclusion of more interactions (even if they are in some 
sense small), breaking thereby the huge symmetry explicitly. Nevertheless, 
the quantization of the volume-preserving diffeomorphisms may be of some 
independent interest, although the solution of this problem might be 
difficult.  
Finally, the semi-classical quantization of the remaining degrees of freedom, 
which are not symmetries, probably just amounts to a renormalization of the 
coupling constants in the effective field theory. These are usually taken 
into account implicitly by fitting the coupling constants to experimentally 
measured quantities.  
\\
In any case, we think that we have identified and solved
an important submodel in the
space of Skyrme-type effective field theories, which is singled out both by its
capacity to reproduce qualitative properties of the liquid droplet
approximation of nuclei, at least at the classical level,  and by its unique 
mathematical structure.
The model directly relates the nuclear mass to the topological charge, and it
naturally provides both a finite size for the nuclei and the liquid droplet
behaviour, which probably is not easy to get from an effective field
theory. So our model solves a conceptual problem by explicitly deriving said
properties from a (simple and solvable) effective field theory.
Last not least, our exact solutions might provide a calibration for the 
demanding
numerical computations in physical applications of more generalized Skyrme
models.
\\
Given this success, it is appropriate to discuss the circumstances  
which make the model relevant.
First of all, from a fundamental QCD point
of view, there is no reason to neglect the sextic term, just like there is no
reason to ignore the quadratic and quartic terms $L_2$ and $L_4$. 
So the good properties of the $L_{06}$ model seem to indicate that in certain
circumstances the sextic term could be more important than the terms $L_2$ and
$L_4$. The quadratic term is kinetic in nature, whereas the
quartic term provides, as a leading behaviour, two-body interactions. On the
other hand, the sextic term is essentially topological in nature, being the
square of the topological current (baryon current). So in circumstances where
our model is successful this seems to indicate that a {\em collective} 
(topological)
contribution to the nucleus is more important than kinetic or two-body
interaction contributions. This behaviour is, in fact, not so surprising for a
system at strong coupling (or for a strongly non-linear system).   
A detailed study of the generalizations mentioned above, or of the more
conceptual cosiderations of this paragraph, is beyond the scope of
this letter and will be presented in future publications.
\\
Finally, let us briefly mention a recent paper \cite{inf-vec}, which appeared
after completion of this letter.  
There, a generalized Skyrme
model saturating a Bogomolny bound is derived along completely different lines.
The model of that paper consists of a Skyrme field coupled to an infinite
tower of vector mesons, where these vector mesons may be interpreted as the
expansion coefficients in a basis of eigenfunctions along a fourth spatial
direction. Simple Yang--Mills theory in four Euclidean dimensions is the 
master theory which gives rise to the generalized Skyrme model
via the expansion into the eigenfunctions along the fourth direction, and the
Bogomolny equation for the latter is a simple consequence of the self-duality
equations for instantons in the former theory. If only a finite number of
vector mesons is kept, the topological bound is no longer saturated, but
already for just the first vector meson, the energies are quite close to their
topological bounds. This latter observation might be in some sense related to
the results for our model, because integrating out the vector meson produces
precisely the sextic Skyrme term in lowest order. One wonders whether it is
possible to integrate out all the vector mesons, which
should lead directly to a topological (Bogomolny) version of the Skyrme 
model.   

\section*{Acknowledgements}

C.A. and J.S.-G. thank MCyT (Spain), FEDER (FPA2005-01963) and
Xunta de Galicia (grant INCITE09.296.035PR and
Conselleria de Educacion) for financial support. 
A.W. acknowledges support from the
Ministry of Science and Higher Education of Poland grant N N202
126735 (2008-2010).  Further, A.W. thanks Prof M.A. Nowak for an interesting 
discussion.

 \end{document}